\documentstyle[12pt,epsf,amssymbols]{article}
\title{Separation of variables for the classical and quantum Neumann model.}
\author{O. Babelon~$^*$ \and M. Talon~\thanks{L.P.T.H.E.  Paris
VI, Tour 16, $1^{er}$ \'etage, 4 place Jussieu,  F-75252 PARIS
CEDEX 05} }
\date{}

\begin{document}
\begin{titlepage}
\renewcommand{\thepage}{}
\maketitle
\vskip 2cm
\begin{abstract}
The method of separation of variables is shown to apply to both the
classical and quantum Neumann model. In the classical
case  this nicely yields the linearization
of the flow on the Jacobian of the spectral curve.
In the quantum case the Schr\"odinger equation separates into one--dimensional
equations belonging to the class of generalized Lam\'e differential equations.
\end{abstract}
\vfill
Work supported by CNRS
\newline
PAR-LPTHE 92-01
\end{titlepage}
\renewcommand{\thepage}{\arabic{page}}

\section{Introduction}

The classical Neumann model describes the motion of a particle
constrained to lie on an $(N-1)$-sphere in $N$ dimensional space and
submitted to harmonic forces~\cite{Neu}. Originally this model has
been solved by Neumann~\cite{Neu} for $N=3$ by separation of variables
in the Hamilton-Jacobi equation, and
subsequently K. Uhlenbeck has shown that the model is Liouville integrable by
exhibiting $(N-1)$ conserved quantities in involution~\cite{Uhl}. The
classical solution of J. Moser~\cite{Mo} was shown to be deeply
related to the theory of
hyperelliptic curves by Mumford~\cite{Mum} emphasizing the {\em algebraic}
integrability of the model.

Recently, new  methods for solving integrable models have
been introduced~\cite{AvM,Ko,Ad,Sy,Skl1}, based on underlying Lie algebraic
structures. The relation between the group-theoretical  and
the algebraic--geometrical approaches has been
explained in~\cite{RSTS2}.

However the old methods of resolution by separation of variables
begin to attract new interest~\cite{Skl4}. As a matter of fact the
solution of the quantum
Toda chain was given in Gutzwiller~\cite{Gu} and
Sklyanin~\cite{Skl2} by a method closely connected to this procedure.

We show in this article that the general Neumann model (any $N$)
admits such a resolution both at the classical and {\em quantum} level. This
is achieved by choosing appropriate coordinates on the sphere,
generalizing Neumann's coordinates for $N=3$~\cite{Neu,Mo}.
Considering what we know
about the Lax formulation of the model~\cite{AT5}, this choice
of coordinates is also
strongly reminiscent of the one advocated by Sklyanin~\cite{Skl4}.
The simplifying feature of this model as compared to the general case,
is that the separation of variables is achieved by a transformation
involving only the space coordinates and not the momenta~\cite{Skl4}.

We first recall how the separation of variables in the Hamilton-Jacobi
equation works in the classical case. This is a straightforward
generalization of the work of Neumann~\cite{Neu,Mo}. One gets
$(N-1)$ one--dimensional ordinary differential equations, which are
readily solved.
We will see that the resolution of the Hamilton--Jacobi equation in
this situation of a Liouville integrable system provides a canonical
transformation which allows to linearize at one stroke the flows of
the complete family of hamiltonians in involution.
The separation of variables naturally lead to Abel
sums and provide a simple  proof of the
linearization of the flow on the Jacobian of the Riemann surface
associated to this dynamical problem by the Lax spectral
theory~\cite{AvM,Mum,AT1}. In the quantum case, the Schr\"odinger equation
similarly separates itself into $(N-1)$ one--dimensional equations.
The quasi--classical quantization conditions are readily obtained.
The implications of the exact quantization conditions will be studied
elsewhere.

\section{The Neumann model}

The Neumann model has two equivalent formulations, one
in terms of constrained dynamical variables with Dirac brackets,
and an unconstrained one with canonical Poisson
brackets. These formulations remain equivalent at the quantum
level~\cite{AT1} so we shall stick to the unconstrained one
which is more convenient for our present purposes.

We start from a $2 N$-dimensional free phase space $\{ x_n, y_n,\quad
n = 1 \cdots N \}$ with canonical Poisson brackets: $\{x_n,y_m\} =
\delta_{nm}$ and introduce the ``angular momentum'' antisymmetric
matrix: $J_{kl} = x_k y_l - x_l y_k$ and the Hamiltonian:
\begin{equation}
H = {1 \over 4} \sum_{k \neq l} J_{kl}^2 ~+ ~
{1 \over 2} \sum_k a_k x_k^2
\label{hamiltonian}
\end{equation}
We shall assume in the following that: $a_1<a_2<\cdots<a_N$.
The hamiltonian equations are, with $X =(x_k) \quad  Y = (y_k)$
and the diagonal constant matrix $A =(a_k \delta_{kl})$:
$$\dot{X} = - J X \qquad \dot{Y} = -J Y - A X$$
They automatically ensure that $\sum x_k^2$ remains constant
and lead to the non-linear Newton equations for the particle:
$$\stackrel{..}{x}_k = - a_k x_k - x_k \sum_l ( \dot{x_l}^2 - a_l
x_l^2 )$$

Let us note that the above hamiltonian is invariant under the
{\em canonical} transformations:
\begin{equation}
(x_k,y_k) \longrightarrow (x_k,y_k+\phi x_k)
\label{invariance}
\end{equation}
where $\phi$ may be any function of $\sum_k x_k^2$.
Classically the constrained system can be viewed as the hamiltonian
reduction of the above unconstrained one under this symmetry.

The Liouville integrability of this system is a consequence of the
existence of $(N-1)$ independent quantities in involution, first found
by K. Uhlenbeck:
\begin{equation}
F_k = x_k^2 + \sum_{l \neq k} {J_{kl}^2 \over a_k - a_l}
\label{conserved}
\end{equation}
Notice that $H = 1/2 \,\sum_k a_k F_k$ and $\sum_k F_k = 1$.

\section{Neumann's coordinates on $S^{(N-1)}$.}

Generalizing Neumann's choice in~\cite{Neu}, we shall introduce
$(N-1)$ parameters $t_1,\cdots,t_{N-1}$ on the sphere. They are the
roots of the equation:
$$u(t)\equiv \sum_k {x_k^2 \over t - a_k}=0$$
Notice that this equation is invariant by $x_k \longrightarrow \lambda
x_k$ so that $t_1 < t_2 < \cdots <t_{N-1}$ are indeed defined on the
sphere. Conversely, by definition of the $t_j$ we have for $x\in S^{(N-1)}$:
$$u(t)={\prod_j(t-t_j) \over \prod_k(t-a_k)} \quad\Longrightarrow\quad
x_k^2={\prod_j(a_k-t_j)\over \prod_{l\neq k}(a_k-a_l)}.$$

Considering the graph of $u(t)$ it is easy to see that:
$$a_1<t_1<a_2<t_2<a_3<\cdots<t_{N-1}<a_N$$ and we have a bijection of
this domain $\cal D$ of the $t_j$'s on the ``quadrant'' $x_k>0~ \forall k$ of
the sphere. In general the sphere appears as a $2^N$--fold covering of
the domain $\cal D$, ramified on the edges of $\cal D$.
For instance $x_1 = \sqrt{t_1-a_1}~ \varphi(t_2,\cdots,t_{N-1})$ so
that $x_1$ changes its sign when we turn around the ramification point
$t_1=a_1$ in the complex plane. By similar analytic continuations we
can cover the whole sphere.

\proclaim Proposition.
The $(t_j)$'s define an orthogonal system of coordinates on the
sphere.

\noindent A simple proof is obtained by considering, for each root $t_j$ the
vector:
$$\vec{v}_j = \left({x_1\over t_j-a_1},\cdots,{x_N \over
t_j-a_N}\right)$$
We easily check that
$\vec{x}\cdot\vec{v}_j=0,\quad\vec{v}_j\cdot\vec{v}_{j'}=0$ for $j\neq
j'$ and therefore $(\vec{v}_j)$ is an orthogonal basis to the tangent space
to $S^{(N-1)}$ at $\vec{x}$. On the other hand, taking the derivatives
of the equations $u(t_j)=0$ with respect to $t_1,\cdots,t_{N-1}$,
where the $x_k$ are considered as functions of these variables, we
obtain:
$$\vec{v}_i \cdot {\partial\vec{x}\over\partial t_i}=0 \quad \mbox{if~}i\neq
j, \qquad 2 \vec{v}_j\cdot{\partial\vec{x}\over\partial
t_j}=\vec{v}_j^2$$
Therefore:
\begin{equation}
{\partial\vec{x}\over\partial t_j}= {1\over 2}\vec{v}_j.
\label{xsurv}
\end{equation}
As a byproduct, since $\vec{v}_j^2=-u'(t_j)$  we get the metrical tensor:
$$g_{jj'}={\partial\vec{x}\over\partial
t_j}\cdot{\partial\vec{x}\over\partial t_j'} = -{1\over 4}
\delta_{jj'}{\prod_{n\neq j}(t_j-t_n)\over \prod_{k}(t_j-a_k)}.$$

\section{The Hamilton--Jacobi equation}

It is convenient to start from the Lagrangian formalism since this is
well suited to the change of coordinates on the sphere.
$${\cal L}= {1\over 2} \dot{\vec{x}}^2 - {\cal U}=
 {1\over 2} \sum_{jj'} g_{jj'} \dot{t}_j \dot{t}_{j'}  - {\cal U}$$
Considering the polynomial:
$$\prod_k(t-a_k)\;u(t)=\sum_k\;x_k^2\;\prod_{l\neq k}(t-a_l)=
\prod_j(t-t_j)$$
we see that:
$${\cal U} = {1\over 2}\sum_k\;a_k x_k^2={1\over 2}
\left(\sum_k\;a_k -\sum_j\;t_j \right)$$
The conjugate momentum to $t_j$ is: $p_j \equiv \partial{\cal L}/%
\partial\dot{t}_j= g_{jj} \dot{t}_j$ leading to the hamiltonian:
$$H= \sum_j\; p_j \dot{t}_j - {\cal L}={1\over 2}\sum_j\;g^{jj}p_j^2 +
{\cal U}$$
where $g^{jj}=(g^{-1})_{jj}=1/g_{jj}$.
The Hamilton--Jacobi equation is a first--order non--linear partial
differential equation obtained by substituting in $H$: $p_j \to \partial S/%
\partial t_j$. The action $S$ is function of the space coordinates
$t_j$'s. For a fixed energy $E$ the Hamilton--Jacobi equation reads:
\begin{equation}
-2 \;\sum_j\quad{\prod_k(t_j-a_k)\over\prod_{n\neq j}(t_j-t_n)}\;
\left({\partial S\over\partial t_j}\right)^2 + {1\over 2} \left(
\sum_k\;a_k -\sum_j\;t_j \right) - E=0
\label{hamjac}
\end{equation}

The method of separation of variables consists in looking for a
so--called {\em complete} solution of eq.~(\ref{hamjac}), i.e. depending on
$(N-1)$
arbitrary constants, of the form:
$$S(t_1,\cdots,t_{N-1})=S_1(t_1)+\cdots+S_{N-1}(t_{N-1}).$$
For this purpose we need to consider the following Vandermonde
determinant:
$$D\equiv D(t_1,\cdots,t_{N-1})=
\left\vert\matrix{1&\cdots&1\cr
t_1&\cdots&t_{N-1}\cr
\vdots&\ddots&\vdots\cr
t_1^{N-2}&\cdots&t_{N-1}^{N-2}}\right\vert=
\prod_{m> n}(t_m-t_n)$$
We shall also need similar determinants $D_j$ obtained by removing the
$j^{\mbox{th}}$ column and the last row in $D$. One has the useful
identities:
\begin{equation}
{D\over D_j}= (-1)^{N-j-1}\;\prod_{n\neq j}(t_j-t_n)
\label{dsurdj}
\end{equation}

Moreover considering the determinant obtained by replacing the last
row by $t_j^k$ and expanding it over this row, we find:
\begin{equation}
\sum_j\;(-1)^{N-1-j}\;t_j^k D_j =\left\{
\matrix{\bullet&0&k=0,\cdots,N-3\hfill\cr
\bullet&D&k=N-2\hfill\cr
\bullet&(\sum_j\;t_j)\;D&k=N-1\hfill}\right.
\label{magique}
\end{equation}
Finally denoting $\Delta(t)=\prod_k (t-a_k)$ we have in view of~(\ref{dsurdj}):
$$g^{jj}=-4 (-1)^{N-j-1}\,{D_j\over D}\,\Delta(t_j)$$
and we can rewrite equation~(\ref{hamjac}) as:
$$-4 \sum_j\;(-1)^{N-j-1}\;D_j\, \Delta(t_j)\;\left(
{\partial S\over \partial t_j}\right)^2 + D(-\sum_j\;t_j +
\sum_k\;a_k - 2E)=0$$
Using identities~(\ref{magique}) one can separate the variables,
getting:
$$4\; \Delta(t_j)\;\left({\partial S_j\over \partial t_j}\right)^2+
\sum_{k=0}^{N-1}\;c_k \, t_j^k=0$$
where:
$$c_{N-1}=1,\quad c_{N-2}=2E-\sum_k a_k,\quad\mbox{and the other~}c_k
\mbox{~arbitrary.}$$
Let us remark that all $S_j$'s verify the same equation which we can
write as:
\begin{equation}
4\; \Delta(t) \left({dS\over dt}\right)^2+\prod_{n=1}^{N-1}(t-b_n)=0
\label{jacsep}
\end{equation}
where the $b_n$'s are $(N-1)$ independent constants. The energy is
obtained in terms of the $b_n$'s as: $E=1/2\,(\sum_k a_k -\sum_n b_n)$.

\section{The general linear flow on the Liouville torus.}

We have obtained the complete solution of the Hamilton--Jacobi
equation:
$$S(t_1,\cdots,t_{N-1};b_1,\cdots,b_{N-1})=\sum_j\;S_j(t_j;b_1,\cdots,b_{N-1}).$$
One can interpret this action as a {\em generating function} for a canonical
transformation from the original variables $(t_j,p_j)$ to new
variables $(b_n,\psi_n)$ linearizing the equations of motion.
As usual, one has:
$$p_j={\partial S \over \partial t_j},\quad \psi_n={\partial S
\over \partial b_n}$$
This transformation is at the level of the symplectic structure and
does not refer to any specific hamiltonian. Since all the $F_k$
in eq.~(\ref{conserved}) play the same role it is natural to consider
the Hamilton--Jacobi equations with respect to any one of the
hamiltonians in involution $F_k$. We show in this section that they all
lead to the {\em same separated equations} for $S$.

In order to do that, we first express $J_{kl}$ in terms of  the
coordinates $t_j$ and their conjugate momenta $p_j$. Consider the
canonical 1--form $\alpha= \sum_k y_k dx_k $ which reduces on the
sphere to $\alpha = \sum_j p_j dt_j$. So we have in view of eq.~(\ref{xsurv})
$$p_j=\sum_k \;y_k {\partial x_k \over \partial t_j}\quad
\mbox{whence} \quad \vec{y}\cdot \vec{v}_j=2 \,p_j~.$$
This determines $\vec{y}$ modulo a vector proportional to $\vec{x}$ which does
not
affect the value of $J_{kl}$. We find $\vec{y}= 1/2\,\sum_j \,
g^{jj} p_j \vec{v}_j$ and:
$$J_{kl}= -{1\over 2}\,\sum_j\,(a_k-a_l)\,v_j^kv_j^l\,g^{jj}p_j$$
With the help of this formula, $F_k$ may be expressed as:
$$F_k=x_k^2\left(1-\sum_j{g^{jj}\,p_j^2 \over t_j-a_k}\right)$$

It is convenient~\cite{AT5} to introduce the generating function for
the $F_k$:
\begin{equation}
{\cal H}(\lambda)\equiv \sum_k {F_k \over \lambda -a_k} = {\prod_n
(\lambda - b_n)\over \prod_k(\lambda - a_k)}
\label{hdel}
\end{equation}
for appropriate $b_n$'s and we have used $\sum F_k =1$.
By a simple calculation we find:
\begin{equation}
{\cal H}(\lambda)= u(\lambda)\,\left(1-\sum_j \,{g^{jj}p_j^2
\over t_j-\lambda}\right)
\label{hlclas}
\end{equation}
where $u(\lambda)$ has been introduced in section 3.
Let us consider the Hamilton--Jacobi equation associated to the
hamiltonian ${\cal H}(\lambda)$. It reads:
\begin{equation}
\sum_j \left\{g^{jj}\left({\partial S\over\partial t_j}\right)^2
{1\over t_j-\lambda}\right\}
+{\prod_n(\lambda-b_n)\over \prod_j(\lambda-t_j)} - 1 =0
\label{jachl}
\end{equation}

\proclaim Proposition.
For any $k=0,\cdots,N-2$ we have:
\begin{equation}
\sum_j (-1)^{N-j-1}\;{D_j t_j^k \over t_j-\lambda} = (-1)^N {D
\lambda^k \over\prod_j(t_j-\lambda)}
\label{plusmagique}
\end{equation}

\noindent This is proven similarly to eq.~(\ref{magique}) by considering the
determinant obtained by replacing the last row of $D$
by $t_j^k/(t_j-\lambda)$.
Using now the identity~(\ref{plusmagique}) for $k=0$
one can rewrite eq.~(\ref{jachl}) in the form:
$$\sum_j\;(-1)^{N-1-j}{D_j\over D}{1\over t_j-\lambda}\left[
4\;\Delta(t_j)\,\left({\partial S\over\partial t_j}\right)^2
+ t_j^{N-2}\,(t_j-\lambda)+\prod_n(\lambda-b_n)\right]=0$$
This leads to the separated equations:
$$4\;\Delta(t_j)\,\left({\partial S\over\partial t_j}\right)^2+
\prod_n(\lambda-b_n)+(t_j-\lambda)\left(t_j^{N-2}+\sum_{l=0}^{N-3}
\,c_l(\lambda)t_j^l\right)=0$$
We look for a solution of the form: $S=\sum\;S_j(t_j)$ {\em
independent} of $\lambda$. This is indeed possible and is achieved by
setting:
$$t^{N-2}+\sum_{l=0}^{N-3}\,c_l(\lambda)t^l={
\prod_n(t-b_n)-\prod_n(\lambda-b_n)\over t-\lambda}$$
and we recover exactly eq.~(\ref{jacsep}). In addition we have
obtained an interpretation of the $b_n$'s in eq.~(\ref{jacsep}):
they are the
roots of ${\cal H}(\lambda)$. So we have the expression of K.
Uhlenbeck's conserved quantities $F_k$ in terms of the integration
constants $b_n$ of the Hamilton--Jacobi equation:
$$F_k = {\prod_n(a_k-b_n)\over\prod_{l\neq k}(a_k-a_l)}$$

We have shown that the $b_n$'s are equivalent to the $F_k$'s as action
variables. Since the symplectic form is:
$$\omega = d \alpha = \sum_j dp_j \wedge dt_j = \sum_n db_n \wedge
d\psi_n$$ we see that the $b_n$'s are in involution. This is still
another proof of the involution property of the $F_k$'s.
Moreover their conjugate variables $\psi_n$'s have a linear time
evolution under ${\cal H}(\lambda)$ for any $\lambda$.
We have proven:
\proclaim Proposition.
The complete solution of the family of Hamilton--Jacobi equations
${\cal H}(\lambda)$ is $\sum_j S(t_j)$ for $S$ solution
of~(\ref{jacsep}) and  ${\cal H}(\lambda)$ is expressed in terms of
the $b_n$'s only.

\section{Discussion of the classical solution.}

As an application let us solve the equations of motion for the
coordinates $t_k$ for any hamiltonian ${\cal H}(\lambda)$.
These equations are readily obtained:
$$\dot{t}_j =\{{\cal H}(\lambda),t_j\}=-2\,u(\lambda)\;{
g^{jj}\,p_j \over t_j-\lambda}\quad\mbox{with}\quad p_j\to
{\partial S\over \partial t_j}\;.$$
Introducing the polynomial:
$$P(t)= \prod_{k=1}^N(t-a_k)\prod_{n=1}^{N-1}(t-b_n)$$
and denoting by $\tau$ the time variable we get:
\begin{equation}
{dt_j\over\sqrt{-P(t_j)}}= {4 u(\lambda) d\tau \over
\prod_{n\neq j}(t_j-t_n)\,(t_j-\lambda)}
\label{motion}
\end{equation}

Now it is natural to introduce the hyperelliptic Riemann
surface of genus $(N-1)$ defined in the complexified plane
$\Bbb{C}^2$ by the equation:
\begin{equation}
s^2+P(t)=0
\label{courbe}
\end{equation}
As a matter of fact this curve also appears as the spectral curve,
in the Lax pair approach~\cite{AT5}. The independent
abelian differentials of first class on this surface are the $t^k \,dt/s$
for $k=0,\cdots,N-2$.
Recalling the identity~(\ref{plusmagique}) we get, for precisely these
values of $k$:
\begin{equation}
\sum_j\;{t_j^k \,dt_j\over s_j}=4\,d\tau\,u(\lambda)\,
\sum_j\;(-1)^{N-1-j}{t_j^k\over t_j-\lambda}{D_j\over D}=
-4\,{\lambda^k\over \prod_l\,(\lambda -a_l)}\,d\tau
\label{abelsum}
\end{equation}

Let us interpret geometrically this remarkable relation: to the point
$\vec{x}$ on the sphere are associated $(N-1)$ unordered
points $(t_j,s_j)$ on
the Riemann surface of equation~(\ref{courbe}), the $t_j$'s being the
roots of $u(t)=0$ and $s_j = \sqrt{-P(t_j)}$.
Such an {\em unordered} set  defines a
{\em divisor\/} on the surface, and since $(N-1)$ is the
genus of this surface  the path integrals of the Abel sums appearing in the
left--hand side of equation~(\ref{abelsum}) provide an analytic
bijection into the Jacobian torus of the Riemann curve. The right--hand
side then shows that this point on the Jacobian evolves linearly with
time. Of course the physical quantities must be real, so the divisor
moves on a connected component of the real slice of the Jacobian,
see~\cite{Mum},
which can be identified with the Liouville torus.

Let us illustrate this geometry for  Neumann's case of $N=3$. The
Riemann surface is of genus 2, and may be pictured as:
\[
\epsffile[36 72 468 252]{surface.ps}
\]
There are two abelian differentials: $dt/s$ and $t\, dt/s$ leading to
two path integrals ($P_0$ is any origin):
\begin{eqnarray*}
\Omega_1 &=& \int_{P_0}^{(t_1,s_1)} \,{dt\over s} +
\int_{P_0}^{(t_2,s_2)} \,{dt\over s}\\
\Omega_2 &=& \int_{P_0}^{(t_1,s_1)} \,{t\, dt\over s} +
\int_{P_0}^{(t_2,s_2)} \,{t\, dt\over s}
\end{eqnarray*}
Obviously $\vec{\Omega}=(\Omega_1,\Omega_2)$ has 4 periods
corresponding to loops around a homology basis of the Riemann surface.
The application $((t_1,s_1),(t_2,s_2))\longrightarrow\vec{\Omega}$ maps
the Jacobian of the surface (analytic space of divisors modulo
equivalence) to a 2--dimensional complex torus.

Finally we notice that the action itself is given by a similar
integral:
$$S(t_1,\cdots,t_{N-1})={1\over 2}\sum_j \int_{P_0}^{P_j=(t_j,s_j)}
\,{\prod_n(t-b_n)\over s}\,dt$$
Remark that the terms $t^k$ for $k=0,\cdots,N-2$ in $\prod_n(t-b_n)
=t^{N-1}+\sum_0^{N-2}\,c_k\,t^k$
lead to abelian integrals of first kind, while the term $t^{N-1}$
leads to an integral of second kind, having a double pole at $\infty$. So the
action may be seen as a multivalued meromorphic function on the
Jacobian. More precisely, with similar notations, one can write:
$$S={1\over 2}\sum_{k=1}^{N-1}\,c_{k-1}\,\Omega_k+S_0$$
where the $c_k$'s are linear functions of $F_k$'s with coefficients
polynomials in the $a_k$'s, since $\prod_n(t-b_n)=\prod_k(t-a_k)\sum_k\,
F_k/(t-a_k)$ and $S_0$ is:
$$S_0(P_1+\cdots+P_{N-1})={1\over 2}\,\sum_j\int_{P_0}^{P_j}\,{t^{N-1}
\,dt \over s}\;.$$
As a function of the divisor $P_1+\cdots+P_{N-1}$, $S_0$
has a simple pole on a variety of codimension 1 obtained when one of
the $P_j$'s goes to $\infty$, which is well known to be the divisor of
a theta function on the Jacobian torus~\cite{Mum}. By restriction to
real variables, $S$ becomes a real multivalued {\em analytic} function
on the $(N-1)$--dimensional real Liouville torus.

\section{The quantum case}

The quantization of the Neumann system will be performed in the
unconstrained formalism. The constraint will appear at the level of
the wave functions. We consider $2N$ self--adjoint operators
$\{x_n,y_n,~n=1,\cdots,N\}$ with canonical commutation relations:
$$[x_k,x_l]=[y_k,y_l]=0,\; [x_k,y_l]=i\hbar \delta_{kl}\;.$$
We  define $J_{kl}=x_k\,y_l-x_l\, y_k$ so that $J_{kk}=0$.
Notice that there is no ordering ambiguity for $k\neq l$.
Finally the quantum analog of K. Uhlenbeck's operators:
$$F_k = x_k^2 + \sum_{l \neq k} {J_{kl}^2 \over a_k - a_l}$$
are also defined unambiguously and are self--adjoint.

\proclaim Proposition.
The quantum theory is formally integrable, i.e.
$$[\, F_k\, , \, F_l \,]=0$$

\noindent{\bf Proof.}
A proof of this fact was given in~\cite{AT5} using the $R$-matrix
formalism. Since this is a fundamental result we give here a more
direct proof. One  checks immediately that the $J_{kl}$ and $x_k$
obey the algebra, similar to the classical Poisson algebra:
$$[J_{ij},J_{kl}]=i\hbar\left(
\delta_{jk}J_{li}+\delta_{il}J_{kj}+\delta_{jl}J_{ik}+\delta_{ik}J_{jl}
\right)$$
$$[J_{ij},x_k]=i\hbar\left(\delta_{ki}x_j-\delta_{kj}x_i\right)$$
First one then shows that:
$$\left[\sum_{p \neq k} {J_{kp}^2 \over a_k - a_p}\;,\;
\sum_{q \neq l} {J_{lq}^2 \over a_l - a_q}\right]=0$$
We can assume $k\neq l$. Therefore in the commutator $[J_{kp},J_{lq}]$
we have three possibilities; $p=q$, $q=k$, and $p=l$. The case $p=q$
produces the term:
$$\sum_{p\neq k,l}{1\over (a_k -a_p) (a_l -a_p)}[J_{kp}J_{kl}J_{lp}+
J_{kp}J_{lp}J_{kl} + J_{kl}J_{lp}J_{kp} + J_{lp}J_{kl}J_{kp} ]$$
The case $q=k$ gives:
$$\sum_{p\neq l}-{1\over(a_k -a_p) (a_k-a_l)}[J_{kp}J_{lp}J_{lk}+
J_{kp}J_{lk}J_{lp} + J_{lp}J_{lk}J_{kp} + J_{lk}J_{lp}J_{kp} ]$$
Notice that this term vanishes for $p=k$ so that one can assume
$p\neq k,l$ in the sum over $p$. Finally the case $p=l$
similarly produces (with $q \to p$):
$$\sum_{p\neq l,k} {1\over (a_l -a_p) (a_k -a_l)}[J_{kl}J_{pk}J_{lp}
+ J_{kl} J_{lp} J_{pk} + J_{pk}J_{lp}J_{kl}+ J_{lp}J_{pk}J_{kl}]$$
In this last expression,  in the first and last terms
one can bring $J_{kl}$ in the middle position.
In doing so the commutators cancel. Then  the three terms
have the same $J$ factors and add to  zero.
Similarly one shows that:
$$\left[x_k^2\, ,\, \sum_{q \neq l} {J_{lq}^2 \over a_l - a_q}\right]-
\left[x_l^2\, ,\, \sum_{p \neq k} {J_{kp}^2 \over a_k - a_p}\right]=0$$
so that finally $[F_k,F_l]=0$.

The canonical commutation relations are realized by setting:
$$y_k=-i\hbar{\partial\over\partial x_k}\;.$$
We look for a simultaneous diagonalization of the $F_k$'s, i.e.:~
$F_k\,\Psi=f_k\,\Psi\,$.
As in the classical case they are invariant under the
symmetry~(\ref{invariance}). This symmetry translates on the wave
function by:
$$\Psi \longrightarrow e^{i \varphi(r)}\,\Psi\; .$$
Here $\varphi$ is simply related to $\phi$ by $\phi=\hbar/r\;\varphi'(r)$
, and $r^2=\sum_k\,x_k^2$.
This shows that the symmetry is implemented by restricting oneself to
wave functions independent of $r$. Consequently we will look for a wave
function $\Psi$ depending on Neumann's variables $t_j$ defined on
$S^{(N-1)}$.

As a preparation to the general calculation, we consider first the
Schr\"odinger equation associated to Neumann's hamiltonian:
$H=1/2\,\sum_k\,a_k F_k$. We find:
$$H={1\over 4}\sum_{k,l}\,J_{kl}^2 + {1\over 2}\sum_k\,a_k \,x_k^2.$$
It is easy to compute:
$$\sum_{k,l}\,J_{kl}^2= -\hbar^2 \left(\,2r^2 \,\vec{\nabla}^2 +(4-2
N)\,\vec{x}.\vec{\nabla} -2\,(\vec{x}.\vec{\nabla})^2\,\right)$$
For wave functions independent of $r$ we have
$\vec{x}.\vec{\nabla}\,\Psi=0$ and $r^2\, \vec{\nabla}^2\,\Psi$ reduces to
the orbital part of the Laplacian, i.e. the Laplacian on the sphere.
So we can write without further ado the Schr\"odinger equation:
\begin{equation}
\left({-\hbar^2\over 2}\triangle + {\cal U}\right)\Psi = E \Psi
\label{sch}
\end{equation}
and express the Laplacian on the sphere using Neumann's coordinates as:
\begin{equation}
\triangle={1\over \sqrt{g}}\sum_{jj'}{\partial\over\partial t_j}\;\left(\,
\sqrt{g}\,g^{jj'}\,{\partial\over\partial t_{j'}}\,\right)
\label{lap}
\end{equation}
Here $g=\prod_j g_{jj}$ is the determinant of the metrical tensor.
It involves $\prod_j\,\prod_{n\neq j}(t_j-t_n) =\pm
\prod_{m>n}(t_m-t_n)^2= \pm D^2$ so neglecting numerical factors that
cancel themselves between $\sqrt{g}$ and $1/\sqrt{g}$
we can set $\sqrt{g}=D/\sqrt{\prod_n\Delta(t_n)}>0$.
{}From this we get:
$$\sqrt{g}\,g^{jj}=-4 (-1)^{N-j-1}\; {D_j\over\sqrt{\prod_{n\neq j}
\Delta(t_n)}}\; \sqrt{\Delta(t_j)}$$
and we remark that ${D_j/\sqrt{\prod_{n\neq j}
\Delta(t_n)}}$  is independent of
$t_j$ and can be commuted with $\partial/\partial t_j$  to the
left of $\triangle$.

Finally the equation~(\ref{sch}) reduces to the following simple form:
$$\sum_{j=1}^{N-1}\;(-1)^{N-j-1}\;{D_j\over D}\;\left\{
4\hbar^2\,\sqrt{\Delta(t_j)}\,{\partial\over\partial t_j}\,
\sqrt{\Delta(t_j)}\,{\partial\over\partial t_j}-
t_j^{N-1}+t_j^{N-2}\,(\sum_k a_k-2E)\right\}\,\Psi=0.$$
It is now easy to separate the variables, i.e. to look for a solution
of the form:
$$\Psi(t_1,\cdots,t_{N-1})=\Psi_1(t_1)\,\Psi_2(t_2)\cdots\Psi_{N-1}(t_{N-1})$$
and we find that the $\Psi_j$'s all obey the same ordinary differential
equation:
$$\left\{4\hbar^2\,\sqrt{\Delta(t)}\,{d\over d t}\,
\sqrt{\Delta(t)}\,{d\over d t}-
t^{N-1}+t^{N-2}\,(\sum_k a_k-2E)+\sum_{k=1}^{N-2}\,c_k\,t^{N-k-2}
\right\}\,\Psi(t)=0$$
Writing $t^{N-1}-t^{N-2}\,(\sum_k a_k-2E)-\sum_{k=1}^{N-2}\,c_k\,t^{N-k-2}
=\prod_n(t-b_n)$ and:
$${\prod_n(t-b_n)\over\prod_k(t-a_k)}=\sum_k\,{f_k\over t-a_k}$$
we get the separated Schr\"odinger equation in the form:
\begin{equation}
\left[{d^2\over dt^2}+{1\over 2}\sum_k{1\over t-a_k}\;{d\over dt}
-{1\over 4\hbar^2}\sum_k{f_k\over t-a_k}\right]\,\Psi(t)=0
\label{lame}
\end{equation}
This equation has been studied in the litterature and is known as the
generalized Lam\'e equation~\cite{WW}.

\section{The general Schr\"odinger equation.}

Since all the $F_k$'s commute and are self--adjoint, there exists a
complete set of common eigenvectors. We therefore expect that
similarly to the classical case the same separation of variables occur
for the whole family of hamiltonians ${\cal H}(\lambda)$. We shall
show that it is indeed the case. As a byproduct, this is an
alternative proof that the $F_k$'s commute, and shows that the $f_k$'s
in equation~(\ref{lame}) are their eigenvalues.

For wave functions independent of $r$ we have
by~(\ref{xsurv}), denoting $v_j^k=x_k/(t_j-a_k)$:
$${\partial\over \partial x_k}={1\over 2}\sum_j\,g^{jj}\, v_j^k \,
{\partial\over\partial t_j} \quad\Longrightarrow\quad
J_{kl}={i\hbar\over
2}\sum_j\, (a_k-a_l)\,v_j^k\,v_j^l   \,
g^{jj}\,{\partial\over\partial t_j}\;.$$
Let us compute the generic hamiltonian:
$${\cal H}(\lambda)=\sum_k\,{F_k\over \lambda -a_k}=
u(\lambda)+{1\over 2}\,\sum_{k,l}\,{J_{kl}^2\over (\lambda
-a_k)(\lambda -a_l)}$$
\proclaim Proposition.
We have (compare with the classical formula~(\ref{hlclas})):
$${\cal H}(\lambda)=u(\lambda)\left[\,1-\hbar^2\sum_i
\,{1\over\lambda -t_i}{1\over \sqrt{g}}{\partial\over\partial t_i}\;\left(\,
\sqrt{g}\,g^{ii}\,{\partial\over\partial t_{i}}\,\right)\,\right]$$

\noindent{\bf Proof.}
We only sketch the main steps of the calculation. We first have:
$$\sum_{k,l}\,{J_{kl}^2\over (\lambda -a_k)(\lambda -a_l)}=
-{\hbar^2\over 4}\sum_{k,l,i,j}\; {(a_k-a_l)^2\over
(\lambda -a_k)(\lambda -a_l)}\,v_j^kv_i^k\,g^{jj}
\left[\,v_j^lv_i^l\,{\partial\over\partial t_j}
+2\,v_j^l{\partial v_i^l\over\partial t_j}\,\right]
\,g^{ii}\,{\partial\over\partial t_i}$$
The summation over $k$ can be performed with the help of the following
identities:
\begin{eqnarray*}
\sum_k\,v_j^kv_i^k&=&4\,g_{ii}\delta_{ij}\\
\sum_k\,a_k\,v_j^kv_i^k&=&4\,t_i\,g_{ii}\delta_{ij}\\
\sum_k\,{v_j^kv_i^k\over\lambda -a_k}&=&4\,{g_{ii}\over\lambda -t_i}
\delta_{ij}+{u(\lambda)\over(t_i-\lambda)(t_j-\lambda)}
\end{eqnarray*}
leading to the result:
$${\cal H}(\lambda)=u(\lambda)-{\hbar^2\over 8}\sum_{l,i,j}
\left[{(\lambda -a_l)u(\lambda)\over(t_i-\lambda)(t_j-\lambda)}
+{4g_{ii}(t_i-a_l)^2\delta_{ij}\over(t_i-\lambda)(a_l-\lambda)}\right]
g^{jj}\left[v_j^lv_i^l{\partial\over\partial t_j}
+2 v_j^l{\partial v_i^l\over\partial t_j}\right]
\,g^{ii}\,{\partial\over\partial t_i}$$

To perform the summation over $l$ in addition to the previous
identities we need the following ones:
\begin{eqnarray*}
\sum_l v_j^l{\partial v_i^l\over\partial t_j}&=&
2\,{\partial g_{ii}\over\partial t_i}\delta_{ij}+
2\,{g_{jj}\over t_i-t_j}(1-\delta_{ij})\\
\sum_l a_l\,v_j^l{\partial v_i^l\over\partial t_j}&=&
2\,{\partial (t_i g_{ii})\over\partial t_i}\delta_{ij}+
2\,{t_j g_{jj}\over t_i-t_j}(1-\delta_{ij})\\
\sum_l{1\over\lambda -a_l}v_i^l{\partial v_i^l\over\partial t_i}&=&
{1\over 2}{\partial\over\partial t_i}\left[
{u(\lambda)\over(t_i-\lambda)^2}+4\,{g_{ii}\over\lambda-t_i}\right]
\end{eqnarray*}
The following result then emerges:
$${\cal H}(\lambda)=u(\lambda)\left[\,1+{\hbar^2\over 2}\,\sum_i
\,{1\over\lambda -t_i}\left\{-2\,{\partial\over\partial t_i}
-g^{ii}{\partial g_{ii}\over\partial t_i}+\sum_{j\neq i}
{1\over t_j-t_i}\right\}g^{ii}\,{\partial\over\partial t_i}
\right]$$
The result now follows by noticing that:
$${\partial\log\sqrt{g}\over\partial t_i}=
{1\over 2}\,\left(g^{ii}{\partial g_{ii}\over\partial t_i}-
\sum_{j\neq i} {1\over t_j-t_i}\right)\;.$$

The corresponding Schr\"odinger equation is:
$${\cal H}(\lambda)\,\Psi=\left(\sum_k\,{f_k\over\lambda
-a_k}\right)\,\Psi$$ and defining the $b_n$'s such that
$\prod_n(\lambda -b_n)/\Delta\;(\lambda)=\sum_k\,f_k/(\lambda -a_k)$ it
can be rewritten as:
$$\sum_{j=1}^{N-1}\;{(-1)^{N-j-1}D_j\over D (\lambda -t_j)}\;\left\{
4\hbar^2\,\sqrt{\Delta(t_j)}\,{\partial\over\partial t_j}\,
\sqrt{\Delta(t_j)}\,{\partial\over\partial t_j}-
t_j^{N-2}\,(t_j-\lambda)-\prod_n(\lambda -b_n)\right\}\,\Psi=0$$
As in the classical case this leads to the separated equations:
$$\left\{4\hbar^2\,\sqrt{\Delta(t_j)}\,{\partial\over\partial t_j}\,
\sqrt{\Delta(t_j)}\,{\partial\over\partial t_j}-\prod_n(\lambda -b_n)
-(t_j-\lambda)\left(t_j^{N-2}+\sum_{k=0}^{N-3} c_k(\lambda)
t_j^k\right)\right\}\Psi_j(t_j)=0 \,.$$
We obtain equations independent of $\lambda$ by choosing:
$$t^{N-2}+\sum_{k=0}^{N-3} c_k(\lambda)
t^k={\prod_n(t-b_n)-\prod_n(\lambda -b_n)\over t-\lambda}$$
and we get back eq.~(\ref{lame}) with the $f_k$'s identified with the
eigenvalues of the $F_k$'s.

\section{The semi--classical quantization conditions}

The solution of the quantum Neumann model has been reduced to the
study of the Lam\'e equation~(\ref{lame}). We shall not dwell further
on this question in this paper and content ourselves with a discussion
of the semi--classical quantization. The semi--classical wave function
is of the form:
$$\Psi_{f_1\cdots f_N}(t_1,\cdots,t_{N-1})=\exp\left({i\over\hbar}\,
S(t_1,\cdots,t_{N-1})\right)\;.$$
Notice that the semi--classical quantization conditions:
$$\oint p_k\,dq_k=2\pi\hbar\,n$$ are canonical invariants and mean
that $\Psi$ is {\em univalued} on the Liouville torus since
$\oint p\,dq$ is the variation of $S$ when one describes a
non--trivial loop around the Liouville torus.

In order to understand further these conditions, it is necessary to
describe the corresponding cycles on the curve~(\ref{courbe}).
Since $P(t)=\prod_k(t-a_k)\prod_n(t-b_n)$ one has to worry about the
relative disposition of the $a_k$'s and $b_n$'s which depends on the
signs of the $f_k$'s. This introduces several cases as first noted by
Neumann~\cite{Neu}. In order to simplify the discussion we shall
assume for example that all $f_k$'s are positive, implying:
$$a_1<b_1<a_2<b_2<\cdots<b_{N-1}<a_N$$
as may be seen immediately by considering the graph of $\sum_k
f_k/(t-a_k)$. Moreover we know that:
$$a_1<t_1<a_2<t_2<a_3<\cdots<t_{N-1}<a_N$$
so that classically each $t_j$ is constrained in an interval
$[a_j,b_j]$ or $[b_j,a_{j+1}]$ in view of equation~(\ref{jacsep}) i.e.
each $P(t_j)$ must be $\le 0$. Since $P(t)>0$ for $t>a_N$ one sees
that $t_j\in[b_j,a_{j+1}]$.

Let us forget momentarily $t_1,\cdots,t_{N-1}$ and concentrate on
the motion of $t_j$. Recalling equation~(\ref{motion}) we see that
$dt_j/d\tau$ vanishes when $t_j\to b_j$ and $t_j\to a_{j+1}$ and is of
a fixed sign in this interval. So $t_j$ oscillates between its bounds.
Now when $t_j=a_{j+1}$ in view of section 3, $x_{j+1}=0$ but
$\dot{x}_{j+1}\neq 0$. So the point on the sphere $S^{(N-1)}$ crosses the
limit $x_{j+1}=0$ of its quadrant when $t_j\to a_{j+1}$ and continues
its motion on the symmetric quadrant with respect to the hyperplane
$x_{j+1}=0$. In the space of the $t$ variables this may be seen as an
analytic continuation around the branch point $t_j=a_{j+1}$. To sum up
our discussion one loop on the sphere corresponds to two oscillations
on the interval $[b_j,a_{j+1}]$. Writing that the wave function $\Psi$
resumes its initial value at the end of the loop we get the semi--classical
quantization conditions:
$$4\,\int_{b_j}^{a_{j+1}}{1\over 2}\sqrt{-{\prod_n(t-b_n)\over
\prod_k(t-a_k)}}\,dt=2\pi\hbar n_j\,,\quad n_j\in\Bbb{Z}$$

These $(N-1)$ quantization conditions determine the $b_n$'s or
equivalently the $f_k$'s. Of course the range of the $f_j$'s should be
restricted so that the previously mentioned constraints on the $f_k$'s
be obeyed (including in particular the positivity of the energy for
the case of original Neumann's hamiltonian). Finally let us notice
that the integrals appearing in these conditions express
themselves in terms of some of the periods of abelian integrals on the
Riemann surface~(\ref{courbe}).

\section{Conclusion.}

In this article we have shown that the method of separation of
variables which was known to apply to the classical Neumann model
applies equally well to the quantum Neumann model. This model is
presumably one of the simplest non trivial integrable model which can
be solved in this way. This is because the necessary change of
variables to separate the equations is only a change of coordinates on
the sphere. Moreover we have treated on the same footing the whole
family of integrals of motion in involution discovered by K. Uhlenbeck
in the classical case as well as in the quantum case. This method of
resolution introduces a Riemann surface and the associated Jacobian
torus in a particularly simple and elegant way.

In the classical case this leads to the linearization of the flow on
the Jacobian, while in the quantum case the quantization conditions
involve the real periods of the underlying Riemann surface. We will
promote our semi--classical analysis to an exact one in a coming work.

Recently Sklyanin has reconsidered this method of separation of
variables in various integrable systems~\cite{Skl4}. It is interesting
to remark that in such system as the Toda chain the separated
equations appear to be finite difference equations (Baxter equation)
while in our case we end up with ordinary differential equations.
We expect this to be related to the fact that in the Lax pair
formulation the Neumann model involves a linear Poisson
structure~\cite{AT5} while in the Toda chain the Poisson structure is
quadratic~\cite{Skl2}.

\bigskip
\noindent{\bf Acknowledgements} We thank M. Dubois-Violette and R. Kerner for
helpful discussions.

\end{document}